\documentclass[12pt,a4paper]{article}
\usepackage{epsfig}
\begin{document}
\title{Probing the gauge bosons $Z'$ and $B'$ from the littlest Higgs model
in the high-energy linear $e^{+}e^{-}$ colliders}

\author{Chong-Xing Yue, Wei Wang, Feng Zhang\\
{\small  Department of Physics, Liaoning Normal University, Dalian
116029, China}\thanks{E-mail:cxyue@lnnu.edu.cn}\\}
\date{\today}

\maketitle
\begin{abstract}

The littlest Higgs (LH) model predicts the existence of the new
gauge bosons $Z'$ and $B'$. We calculate the contributions of
these new particles to the processes
$e^{+}e^{-}\rightarrow\overline{f}f$ with $f=\tau, \mu, b,$ or $c$
and study the possibility of detecting these new particles via
these processes in the future high-energy linear $e^{+}e^{-}$
collider (LC) experiments with $\sqrt{s}=500GeV$ and
$\pounds_{int}=340fb^{-1}$. We find that, with reasonable values
of the parameter preferred by the electroweak precision data, the
possible signals of these new particles  might be detected. The
$Z'$  mass $M_{Z'}$ can be explored up to $2.8TeV$ via the process
$e^{+}e^{-}\rightarrow\overline{b}b\ $ for $0.3\leq c \leq 0.5 $
and the $B'$  mass $M_{B'}$ can be explored up to $1.26TeV$ via
the process $e^{+}e^{-}\rightarrow\overline{l}l\ $ for $0.64\leq
c' \leq 0.73 $.
 \vspace{1cm}


\end{abstract}

\newpage
\noindent{\bf I. Introduction}

\hspace{5mm} The hadron colliders, such as the Tevatron and the
future LHC, are expected to directly probe possible new physics
beyond the standard model(SM) up to a scale of a few $TeV$, while
a high-energy linear $e^{+}e^{-}$  collider (LC) is required to
complement the probe of the new particles with detailed
measurement[1]. Some kinds of new physics predict the existence of
new particles that will be manifested as a rather spectacular
resonance in the LC experiments if the achievable
center-of-mass(c.m.) energy $\sqrt{s}$ is sufficient. Even if
their masses exceed the c.m. energy $\sqrt{s}$, the LC experiments
also retain an indirect sensitivity through a precision study of
their virtual corrections to electroweak observables. Thus, a
future LC, such as the Giga $Z$ option of the LC, will offer an
excellent opportunity to study new physics with uniquely high
precision.

Little Higgs models[2, 3, 4] were recently proposed as a kind of
models of electroweak symmetry breaking(EWSB), which can be
regarded as one of the important candidates of the new physics
beyond the SM. Little Higgs models employ an extended set of
global gauge symmetries in order to avoid the one-loop quadratic
divergences and thus provide a new approach to solve the hierarchy
between the $TeV$ scale of possible new physics and the
electroweak scale, $v=246GeV=(\sqrt{2}G_{F})^{-\frac{1}{2}}$. In
these models, at least two interactions are needed to explicitly
break all of the global symmetries, which forbid quadratic
divergences in Higgs mass at one-loop level. EWSB is triggered by
the Colemean-Weinberg Potential, which is generated by integrating
out the heavy degrees of freedom. In this kind of models, the
Higgs boson is a pseudo-Goldstone boson of a global symmetry which
is spontaneously broken at some high scale $f$ by an vacuum
expectation value($vev$) and thus is naturally light. A general
feature of this kind of models is that the cancellation of the
quadratic divergences is realized between particles of the same
statistics.

Little Higgs models are weakly interaction models, which contain
extra gauge bosons, new scalars and fermions, apart from the SM
particles. These new particles might produce characteristic
signatures at the present and future collider experiments[5, 6, 7,
8, 9, 10]. Since the extra gauge bosons can mix with the SM gauge
bosons $W$ and $Z$, the masses of the SM gauge bosons $W$ and $Z$
and their couplings to the SM particles are modified from those in
the SM at the order of $v^{2}/f^{2}$. Thus, the precision
measurement data can give severe constraints on this kind of
models[5, 11, 12, 13, 14, 15, 16, 17].

In general, the new gauge bosons are heavier than the current
experimental limits on direct searches. However, these new
particles may have effects at low energy by contributing to higher
dimension operators in the SM after integrating them out, which
might generate observable signals in the present or future
experiments. For example, Ref.[17] has shown that $Z^{'}$ exchange
and $B^{'}$ exchange can give correction effects on the
four-fermion interactions in the context of the little Higgs
models. In this paper, we will discuss the possibility of
detecting the new neutral gauge bosons $Z^{'}$ and $B^{'}$ in the
future LC experiments with the c.m. energy $\sqrt{s}=500GeV$ and
the integrating luminosity $\pounds_{int}= 340fb^{-1}$ and both
beams polarized[1] via considering their contributions to the
processes $e^{+}e^{-}\rightarrow \overline{f}f$ with $f=\tau, \mu,
b$ and $c$ in the context of the littlest Higgs(LH) model[2]. We
find that these new gauge bosons can indeed produce significant
contributions to these processes in wide range of the parameter
space [$0\leq c \leq 0.5, 0.62\leq c' \leq 0.73$] preferred by the
electroweak precision data. The new gauge bosons $Z^{'}$ and
$B^{'}$ might be observable in the future LC experiments.

The LH model has all essential features of the little Higgs
models. So, in the rest of this paper, we give our results in
detail in the framework of the LH model, although many
alternatives have been proposed[3, 4]. Section II contains a short
summary of the masses and the relevant couplings of the new gauge
bosons $Z^{'}$ and $B^{'}$ to ordinary particles. The total decay
widths of these new gauge bosons are also estimated. Section III
is devoted to the calculation and analysis the relative
corrections of the new gauge bosons $Z^{'}$ and $B^{'}$ to the
cross sections of the processes $e^{+}e^{-}\rightarrow
\overline{f}f$ with $f=\tau,\ \mu,\  b$ and $c$. In Section IV, we
proceed to a comparison of the discovery potential of each process
in the future LC experiment with $\sqrt{s}=500GeV$  and
$\pounds_{int}=340fb^{-1}$. Our conclusions and discussions are
given in Section V.

\noindent{\bf II. The masses and the relevant couplings of the
gauge bosons $Z^{'}$ and $B^{'}$}

\hspace{5mm}As the simplest realization of the little Higgs idea,
the LH model $[2]$ consists of a non-linear $\sigma$ model with a
global $SU(5)$ symmetry which is broken down to $SO(5)$ by a
vacuum condensate $f$ $\sim\Lambda_{s}/4\pi \sim TeV$, which
results in fourteen massless Goldstone bosons. Four of these
massless Goldstone bosons are eaten by the SM gauge bosons, so
that the locally gauged symmetry $SU(2)_{1}\times U(1)_{1}\times
SU(2)_{2}\times U(1)_{2}$ is broken down to its diagonal subgroup
$SU(2)\times U(1)$, identified as the $SM$ electroweak gauge
group. The remaining ten Goldstone bosons transform under the $SM$
gauge group as a doublet H and a triplet $\Phi$. This breaking
scenario also gives rise to the new gauge bosons, such as $Z^{'}$
and $B^{'}$. The masses of the neutral gauge bosons $Z^{'}$ and
$B^{'}$ can be written at the order of $v^{2}/f^{2}$[5]:
\begin{eqnarray}
M_{Z'}^{2}&=&M_{Z}^{2}C_{W}^{2}[\frac{f^{2}}{s^{2}c^{2}v^{2}}-1
-\frac{5S_{W}^{3}}{2C_{W}}\frac{sc(c^{2}s'^{2}+s^{2}c'^{2})}{s'c'(5C_{W}^{2}s'^{2}c'^{2}
-S_{W}^{2}s^{2}c^{2})}],\\
M_{B'}^{2}&=&M_{Z}^{2}S_{W}^{2}[\frac{f^{2}}{5s'^{2}c'^{2}v^{2}}-1
+\frac{5C_{W}^{3}}{8S_{W}}\frac{s'c'(c^{2}s'^{2}
+s^{2}c'^{2})}{sc(5C_{W}^{2}s'^{2}c'^{2}-S_{W}^{2}s^{2}c^{2})}],
\end{eqnarray}
where $M_{Z}$ is the $Z$ mass predicted by the SM and $f$ is the
scale parameter. Using the mixing parameters $c(s=\sqrt{1-c^{2}})$
and $c'(s'=\sqrt{1-c'^{2}})$, we can represent the SM gauge
coupling constants as $g=g_{1}s=g_{2}c$ and
$g'=g'_{1}s'=g'_{2}c'$. $S_{W}=\sin\theta_{W}$, $\theta_{W}$ is
the Weinberg angle. From Eq.(1) and Eq.(2) we can see that the
values of $M_{Z'}$ and $M_{B'}$ are mainly dependent on the value
of the scale parameters $(f, c)$, and $(f, c')$, respectively. In
general, the gauge boson $B'$ is substantially lighter than the
gauge boson $Z'$. Considering the constraints of the electroweak
precision data on the free parameters $f,\  c,$ and $c'$ in the LH
model, the value of the ratio $ M_{B'}^{2} / M_{Z'}^{2}$ can be
further reduced[8].

 In the LH model, the couplings of the neutral gauge bosons $Z$,
$Z^{'}$ and $B^{'}$ to fermions can be written as:
\begin{eqnarray}
&&e[(g_{L}^{f}+\delta
g_{L}^{f})\overline{f_{L}}\gamma^{\mu}f_{L}+(g_{R}^{f}+\delta
g_{R}^{f})\overline{f_{R}}\gamma^{\mu}f_{R}]Z_{\mu}\nonumber\\
&&+e[g_{L}^{Zf}\overline{f_{L}}\gamma^{\mu}f_{L}+g_{R}^{Zf}\overline{f_{R}}
\gamma^{\mu}f_{R}]Z'_{\mu}
+e[g_{L}^{Bf}\overline{f_{L}}\gamma^{\mu}f_{L}+g_{R}^{Bf}\overline{f_{R}}
\gamma^{\mu}f_{R}]B'_{\mu}
\end{eqnarray}
with
\begin{equation}
g_{L}^{f}=\frac{1}{S_{W}C_{W}}(I_{3}^{f}-Q_{f}S_{W}^{2}),\hspace{0.5cm}
g_{R}^{f}=\frac{1}{S_{W}C_{W}}(-Q_{f}S_{W}^{2}).
\end{equation}
Where $I_{3}^{f}$ is the third component of fermion isospin  and
$Q_{f}$ is the electric charge of fermion $f$ in units of the
position charge $e$. The $\delta g_{L}^{f}$ and $\delta g_{R}^{f}$
represent the correction terms of the tree-level $Z\overline{f}f$
couplings $g_{L}^{f}$\ and \ $g_{R}^{f}$, which come from the
mixing between the gauge boson $Z'$ and the SM gauge boson $Z$.
The general forms of these terms have been given in Ref.[5]. The
relevant forms, which are related the processes $e^{+}e^{-}
\rightarrow$ $\overline{f}f$($f=l$, $b$ and $c$) can be written
as:
\begin{eqnarray}
g_{R}^{Zf}&=&0,\hspace{1cm}g_{L}^{Zu}=-g_{L}^{Zd}=-g_{L}^{Zl}=\frac{1}{S_{W}}\frac{c}{2s},\\
g_{L}^{Bl}&=&\frac{1}{C_{W}}\frac{1}{2s'c'}(c'^{2}-\frac{2}{5}),\hspace{1cm}g_{R}^{Bl}=
\frac{1}{C_{W}}\frac{1}{s'c'}(c'^{2}-\frac{2}{5}),\\
g_{L}^{Bb}&=&-\frac{1}{C_{W}}\frac{1}{6s'c'}(c'^{2}-\frac{2}{5}),\hspace{1cm}g_{R}^{Bb}=
\frac{1}{C_{W}}\frac{1}{3s'c'}(c'^{2}-\frac{2}{5}),\\
g_{L}^{Bc}&=&-\frac{1}{C_{W}}\frac{1}{6s'c'}(c'^{2}-\frac{2}{5}),\hspace{1cm}
g_{R}^{Bc}=-\frac{1}{C_{W}}\frac{2}{3s'c'}(c'^{2}-\frac{2}{5}),
\end{eqnarray}
where $l=\tau, \mu$ or $e$. The couplings of the gauge boson $B'$
to fermions are quite model dependent, which depend on the choice
of the fermion $ U(1) $ charges under the two $ U(1)$ groups[5,
13]. The $ U(1) $ charges of the SM fermions are constrained  by
requiring that the Yukawa couplings are gauge invariant and
maintaining the usual SM hypercharge assignment. Combing the gauge
invariance of the Yukawa couplings with the $ U(1)$ anomaly-free
can fix all of the $U(1)$ charge values. The couplings of the $ B'
$ with fermions given by Eqs.(6)--(8) come from this kind of
choice. Certainly, this is only one example of all possible $U(1)$
charge assignments. In other little Higgs models, several
alternatives for the $U(1)$ charge choice exist[3, 4, 13].

In the LH model, the custodial $ SU(2) $ global symmetry is
explicitly broken, which can generate large contributions to the
electroweak  observables. If one assumes that the SM fermions are
charged only under $ U(1)_1$, then global fits to the electroweak
precision data produce rather severe constraints on the parameter
space of the LH model[11, 12]. However, if the SM fermions are
charged under $U(1)_{1}\times U(1)_{2}$, the constraints become
relaxed. The scale parameter $f=1\sim 2TeV$ is allowed for the
mixing parameters $c$ and $c'$ in the ranges of $0\sim 0.5$ and
$0.62 \sim 0.73$, respectively[13, 14]. On the other hand, the
neutral gauge boson $B'$ is typically light and should produce
significantly contributions to observables. Thus, it can be seen
as the first signal of the LH model. The production and the
possible signals of the $B'$ at the hadron colliders(Tevatron or
LHC) have been studied in Refs[5, 11, 14]. It has been shown that
the gauge boson $B'$ is excluded for a mass lower than $500GeV$ by
the direct search at the Tevatron. However, Ref.[8] has shown that
a large portion of the parameter space consistent with the
electroweak precision data can accommodate the Tevatron direct
searches to new gauge bosons decaying into dileptons. The light
$B' $ is not excluded by the direct searches for the neutral gauge
boson at the Tevatron. So, we will take the $M_{Z'} $, $M_{B'} $
and the mixing parameters $ c, c' $ as free parameters in our
discussions, which are assumed in the ranges of $1TeV\sim 3TeV$,
$400GeV\sim 1000GeV$, $0.1\sim 0.5$ and $0.62\sim 0.73$,
respectively.

\vspace*{9cm}
\begin{figure}[htb]
\vspace{-10cm}
\begin{center}
\epsfig{file=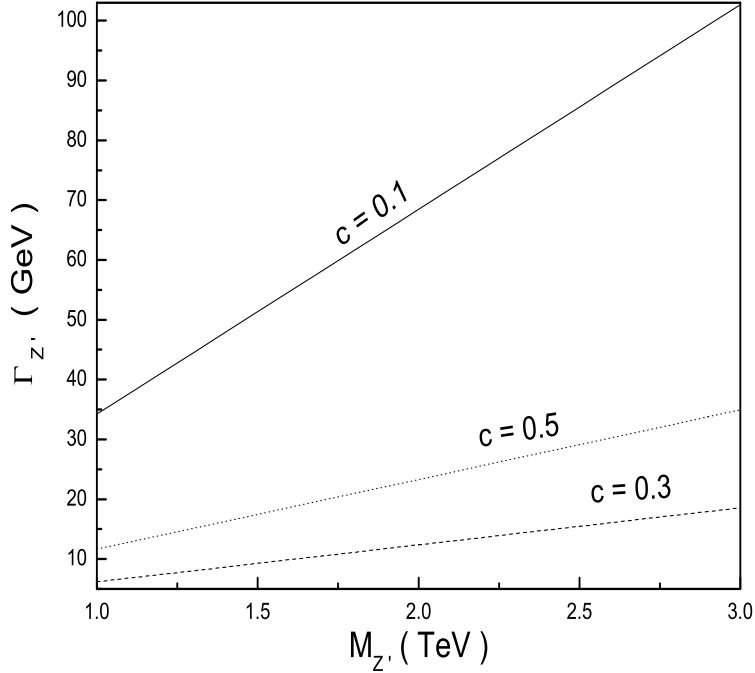,width=320pt,height=300pt} \vspace{-1cm}
\hspace{0.5cm} \caption{The total decay width $\Gamma_{Z'}$ as a
function of the $Z'$ mass $M_{Z'}$ for $c=0.1$(solid
\hspace*{1.8cm}line), $c=0.3$(dashed line) and $c=0.5$(dotted
line).} \label{ee}
\end{center}
\end{figure}

At the leading order, the two-body decay channels of the neutral
gauge boson $V(Z'$ or $B')$ mainly contain
$V\rightarrow\overline{f}f$, where $f$ is any of the SM quarks or
leptons, and $V\rightarrow Zh$. If we ignore the fermion masses,
the generic partial decay width for $V$ to  fermion pair can be
written as [5, 7]:
\begin{equation}
\Gamma(V\rightarrow\overline{f}f)=\frac{C}{24\pi}[(g_{L}^{Vff})^{2}+(g_{R}^{Vff})^{2}]M_{V},
\end{equation}
where $C$ is the fermion color factor and $C=1$(3) for
leptons(quarks). For the gauge boson $Z'$, the total decay width
can be approximately written as:
\begin{eqnarray}
\Gamma_{Z'}&=&6\Gamma(Z'\rightarrow
\overline{q}q)+6\Gamma(Z'\rightarrow \overline{l}l)
+\Gamma(Z'\rightarrow Zh) +\Gamma(Z'\rightarrow W^{+} W^{-}) \nonumber\\
&\approx&
\frac{\alpha_{e}M_{Z'}}{96S_{W}^{2}}[\frac{96c^{2}}{s^{2}}+
\frac{(c^{2}-s^{2})^{2}}{s^{2}c^{2}}].
\end{eqnarray}

In general, the $B'$ mass $M_{B'}$ is not too large and can be
allowed to be in the range of a few hundred $GeV$. So, for the
decay channels $B'\rightarrow\overline{t}t$ and $B'\rightarrow
Zh$, we can not neglect the final state masses. The total decay
width of the gauge boson $B'$ can be written as:
\begin{eqnarray}
\Gamma_{B'}&=&3\Gamma(B'\rightarrow\overline{l}l)
+3\Gamma(B'\rightarrow\overline{\nu}\nu)
+3\Gamma(B'\rightarrow\overline{d}d)\nonumber\\
&&+2\Gamma(B'\rightarrow\overline{u}u) +\Gamma(B'\rightarrow
\overline{t}t)+\Gamma(B'\rightarrow Zh)\nonumber\\
&\approx &
\frac{\alpha_{e}M_{B'}}{4C_{W}^{2}}\{{\frac{85(c'^{2}-\frac{2}{5})^{2}}{18s'^{2}c'^{2}}
+\frac{\sqrt{1-4r_{t}}}{s'^{2}c'^{2}}\{[\frac{5}{6}(\frac{2}{5}-c'^{2})-
\frac{1}{5}x_{L}]^{2}}(1+2r_{t})\nonumber\\
&&+[\frac{1}{2}(\frac{2}{5}-c'^{2})-\frac{1}{5}x_{L}]^{2}(1-4r_{t})\}
+\frac{(c'^{2}-s'^{2})^{2}}{24c'^{2}s'^{2}}\lambda^{\frac{1}{2}}[(1+r_{Z}-r_{h})^{2}+8r_{Z}]\},
\end{eqnarray}
where $r_{i}=m_{i}^{2}/M_{B'}^{2}$ and
$\lambda=1+r_{Z}^{2}+r_{h}^{2}+2r_{Z}+2r_{h}+2r_{Z}r_{h}$. The
mixing parameter between the SM top quark $ t$ and the vector-like
quark $T$ is defined as
$x_{L}=\lambda_{1}^{2}/(\lambda_{1}^{2}+\lambda_{2}^{2})$, in
which $\lambda_{1}$ and $\lambda_{2}$ are the coupling parameters.
In above equation, we have neglected the decay width
$\Gamma(B'\rightarrow W^{+} W^{-})$, which is suppressed by a
factor of $\nu^{4}/f^{4}$.

From above equations, we can see that the total decay width
$\Gamma_{Z'}$ is mainly dependent on the free parameters $M_{Z'}$
and $c$, while the total decay width $\Gamma_{B'}$ is sensitive to
the free parameters $M_{B'}$ and $c'$. For $c' = \sqrt{2/5}$, the
gauge boson $ B' $ mainly decays to $ Zh $ and $ t\overline{t} $,
and the decay modes $l\overline{l}$ and $ q\overline{q}(q\neq t)$
are prohibited being its couplings with the light fermions vanish.
In Fig.1 and Fig.2, we plot $\Gamma_{Z'}$ and $\Gamma_{B'}$ as
functions of $M_{Z'}$ and $M_{B'}$ for three values of $c$ and
$c'$, respectively. In Fig.2 we have taken $\lambda_{1}\approx
\lambda_{2}$. From these figures, we can see that $\Gamma_{Z'}$ is
in tens and up to hundred $GeV$, while $\Gamma_{B'}< 1GeV$ in the
parameter space preferred by the electroweak precision data.

In the following sections, we will use the above formulae to
calculate the corrections of the neutral gauge bosons $Z'$ and
$B'$ to the cross sections of the processes $e^{+}e^{-}\rightarrow
\overline{f}f$ in the parameter space[ $0 \leq c \leq 0.5 $ and
$0.62 \leq c' \leq 0.73 $], which is consistent with the precision
electroweak constraints. Then we will study the realistic
observability of the gauge bosons $Z'$ and $B'$ in the future LC
experiments.

\vspace*{9.5cm}
\begin{figure}[htb]
\vspace*{-10cm}
\begin{center}
\epsfig{file=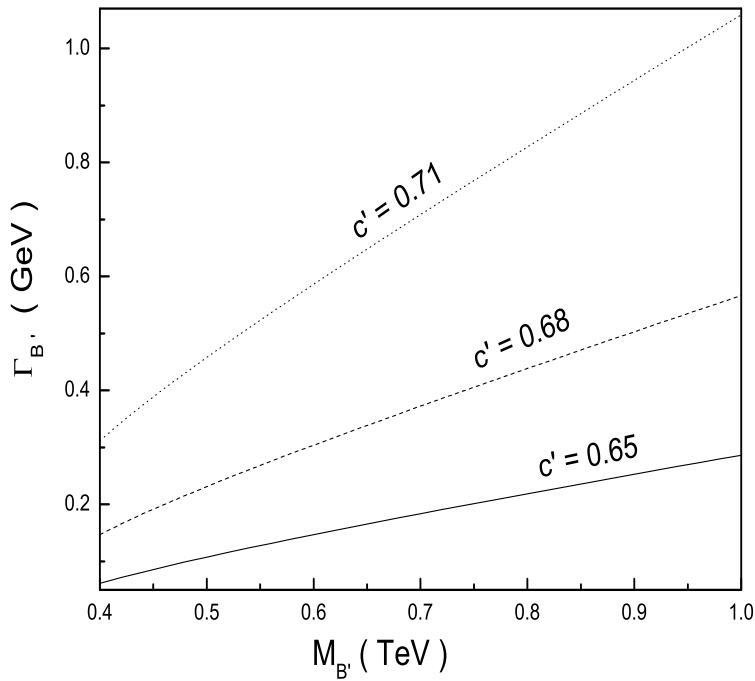,width=320pt,height=300pt} \vspace{-1cm}
\hspace{0.5cm} \caption{The total decay width $\Gamma_{B'}$ as a
function of the $B'$ mass $M_{B'}$ for $c'=0.65$(solid
\hspace*{1.8cm}line), $c'=0.68$(dashed line) and $c'=0.71$(dotted
line).} \label{ee}
\end{center}
\end{figure}

\noindent{\bf III. The contributions of $Z'$ and $B'$ to the
processes $e^{+}e^{-}\rightarrow\overline{f}f$ }

Neglecting fermion mass $m_{f}$($f=\tau,\ \mu,\ b$ or $c$) with
respect to the c.m. energy $\sqrt{s}$, the helicity cross sections
$\sigma_{\alpha \beta}(\overline{f}f)$ of the processes
$e^{+}e^{-}\rightarrow \overline{f}f $ can be given in Born
approximation [18, 19]:
\begin{equation}
\sigma_{\alpha\beta}(\overline{f}f)=N_{c}A|M_{\alpha\beta}(\overline{f}f)|^{2},
\end{equation}
where $\alpha$, $\beta =L,\ R$; $
N_{c}=3(1+\frac{\alpha_{s}}{\pi})$ for quarks and $N_{c}=1$ for
leptons. $A=\sigma(e^{+}e^{-}\rightarrow
r^{*}\rightarrow\overline{f}f)=(4\pi\alpha_{e}^{2})/(3s)$. In the
LH model, the helicity amplitude $M_{\alpha\beta}(\overline{f}f)$
can be written as:
\begin{equation}
M_{\alpha\beta}(\overline{f}f)=Q_{e}Q_{f}+(g_{\alpha}^{e}+\delta
g_{\alpha}^{e})(g_{\beta}^{f}+\delta g_{\beta}^{f})\chi_{Z}
+g_{\alpha}^{Ze}g_{\beta}^{Zf}\chi_{Z'}+g_{\alpha}^{Be}g_{\beta}^{Bf}\chi_{B'}
\end{equation}
with
\begin{equation}
\chi_{i}=\frac{s}{s-M_{i}^{2}+iM_{i}\Gamma_{i}},
\end{equation}
where $\chi_{i}$ represent the propagators of the gauge bosons
$Z,\ Z'$, and $B'$, in which $\Gamma_{i}$ represents the
corresponding total decay width. From above equation, we can see
that the contributions of the LH model to the processes
$e^{+}e^{-}\rightarrow\overline{f}f$ mainly come from three
sources: (1) the modification to the relation between the SM free
parameters, (2) the correction terms to the tree-level $Z
\overline{f}f $ couplings, (3) $Z'$ exchange and $B'$ exchange.

The contributions of the neutral gauge bosons $Z'$ and $B'$ to the
helicity cross sections $\sigma_{\alpha\beta}(\overline{f}f)$ can
be written as:
\begin{eqnarray}
\Delta\sigma_{\alpha\beta}(\overline{f}f)
&=&\sigma_{\alpha\beta}^{ZB}(\overline{f}f)-\sigma_{\alpha\beta}^{SM}
(\overline{f}f)\nonumber\\
& \approx
&2N_{c}A[M_{\alpha\beta}^{SM}(\overline{f}f)g_{\alpha}^{Ze}g_{\beta}^{Zf}\chi_{Z'}
+M_{\alpha
\beta}^{SM}(\overline{f}f)g_{\alpha}^{Be}g_{\beta}^{Bf}\chi_{B'}].
\end{eqnarray}
In above  equation, we have neglected the terms which are
proportional to $(g_{\alpha}^{Bf})^{2}$ and
$(g_{\alpha}^{Zf})^{2}$. This is because the contributions of
these terms to the helicity cross sections are suppressed by the
factor  $\nu^4/f^4 $. The first and second terms of the right-side
of this equation represent the contributions of $Z'$ exchange and
$B'$ exchange, respectively.

The cross sections $\sigma(\overline{f}f)$, which can be directly
detected at the LC experiments, can be written as:
\begin{equation}
\sigma(\overline{f}f)=\frac{1}{4}[\sigma_{LL}(\overline{f}f)+
\sigma_{LR}(\overline{f}f)+\sigma_{RL}
(\overline{f}f)+\sigma_{RR}(\overline{f}f)].
\end{equation}
To discuss the contributions of $Z'$ exchange and $B'$ exchange to
the processes $e^{+}e^{-}\rightarrow\overline{f}f$, we define the
relative correction parameters:
\begin{equation}
R_{1}(\overline{f}f)=\frac{\Delta\sigma_{1}(\overline{f}f)}{\sigma^{SM}(\overline{f}f)},
\hspace{1cm}R_{2}(\overline{f}f)=\frac{\Delta\sigma_{2}(\overline{f}f)}
{\sigma^{SM}(\overline{f}f)},
\end{equation}
where $\Delta\sigma_{1}(\overline{f}f)$ and
$\Delta\sigma_{2}(\overline{f}f)$ represent the contributions of
$Z'$ exchange and $B'$ exchange, respectively. From above
discussions, we can see that $ R_1$ is mainly dependent on the
free parameters $ M_{Z'}$ and $ c $, $ R_2$ is mainly dependent on
the free parameters $ M_{B'}$ and $ c' $.

To see the contributions of the neutral gauge boson $Z'$ to the
processes $e^{+}e^{-}\rightarrow\overline{f}f$, we plot the
relative correction parameters $R_{1}(\overline{l}l)(l=\mu$ or
$\tau)$, $R_{1}(\overline{b}b)$, and $R_{1}(\overline{c}c)$ as
functions of the $Z'$ mass $M_{Z'}$ for three values of the mixing
parameter $c $ in Fig.3, Fig.4, and Fig.5, respectively. From
these figures, we can see that $Z'$ exchange generates negative
contributions to all of these processes. The gauge boson $Z'$ can
decrease the production cross sections of the processes
$e^{+}e^{-}\rightarrow\overline{f}f$. The contributions of $Z'$
exchange to these processes increase as $M_{Z'}$ decreasing and $
c$ increasing. In all of the parameter space of the LH model, the
contributions of $Z'$ to the process
$e^{+}e^{-}\rightarrow\overline{b}b$ are larger than those for the
processes $e^{+}e^{-}\rightarrow\overline{l}l$ or
$e^{+}e^{-}\rightarrow\overline{c}c$. For example, for
$M_{Z'}=1TeV$ and $c=0.5$, the absolute values of the relative
correction parameter $R_{1}$ are $18.2\%$, $11.8\%$, $7.5\%$ for
the processes $e^{+}e^{-}\rightarrow\overline{b}b$,
$\overline{c}c$, and $\overline{l}l$, respectively. For
$M_{Z'}\geq1.7TeV$ and $c\leq0.5$, which satisfy the electroweak
precision constraints[13,14], the absolute value of
$R_{1}(\overline{f}f)$ are smaller than $5\%$, which is very
difficult to be detected in the future LC experiments. However, if
we assume $M_{Z'}\leq 1.5TeV$ and $c > 0.5$, the signal of the
gauge boson $Z'$ can be easy detected.

\begin{figure}[htb]
\vspace*{-0.5cm}
\begin{center}
\epsfig{file=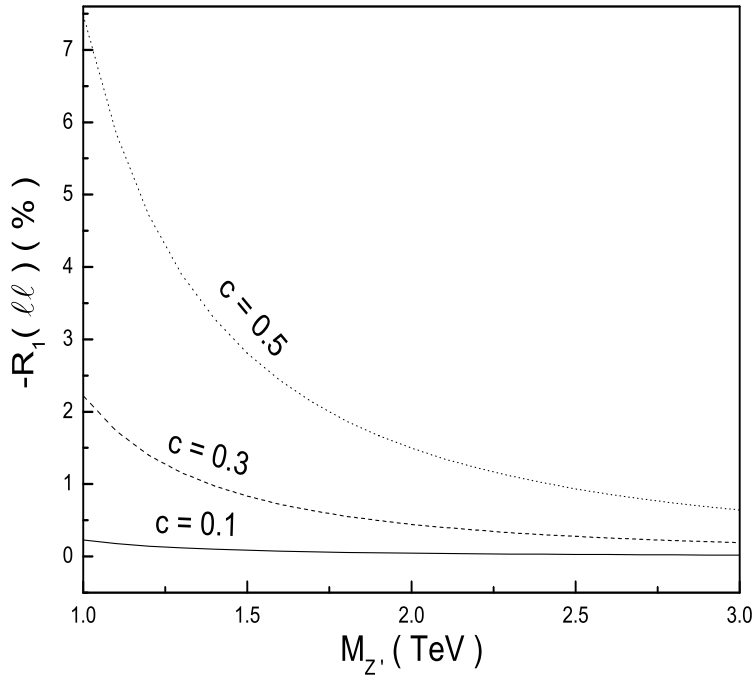,width=320pt,height=300pt} \vspace{-1cm}
\hspace{0.5cm} \caption{The relative correction parameter $R_{1}$
for the process $e^{+}e^{-}\rightarrow \overline{l}l$ as function
\hspace*{1.8cm}of $M_{Z'}$ for $c=0.1$(solid line), $c=0.3$(dashed
line) and $c=0.5$(dotted line).} \label{ee}
\end{center}
\end{figure}

\begin{figure}[htb]
\vspace{-1cm}
\begin{center}
\epsfig{file=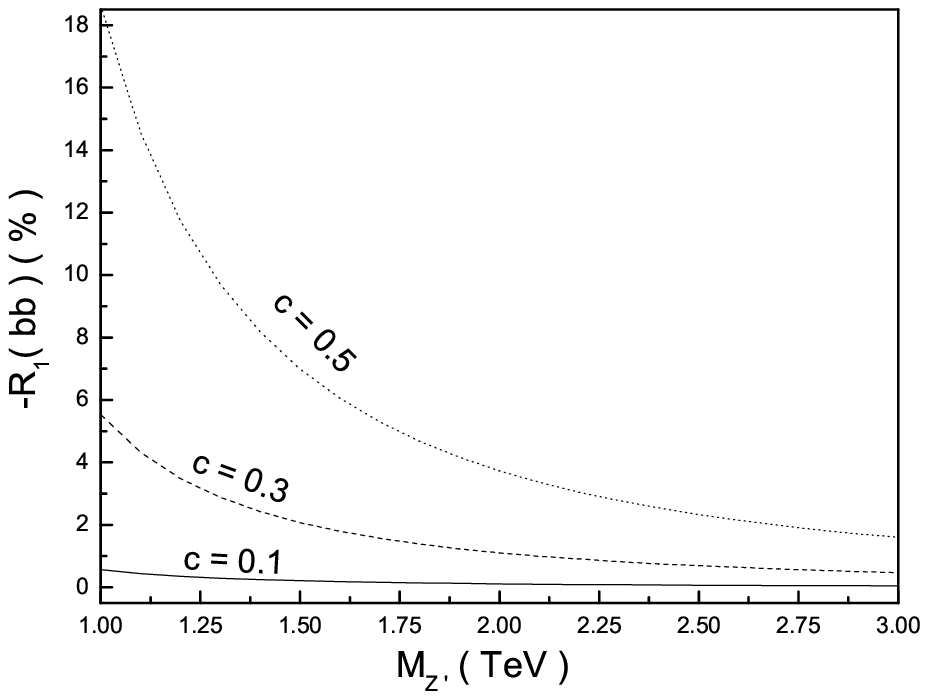,width=320pt,height=300pt} \vspace{-1cm}
\hspace{0.5cm} \caption{Same as Fig.3 but for the process
$e^{+}e^{-}\rightarrow \overline{b}b$.} \label{ee}
\end{center}
\end{figure}
\vspace*{8.5cm}
\begin{figure}[htb]
\vspace{-10cm}
\begin{center}
\epsfig{file=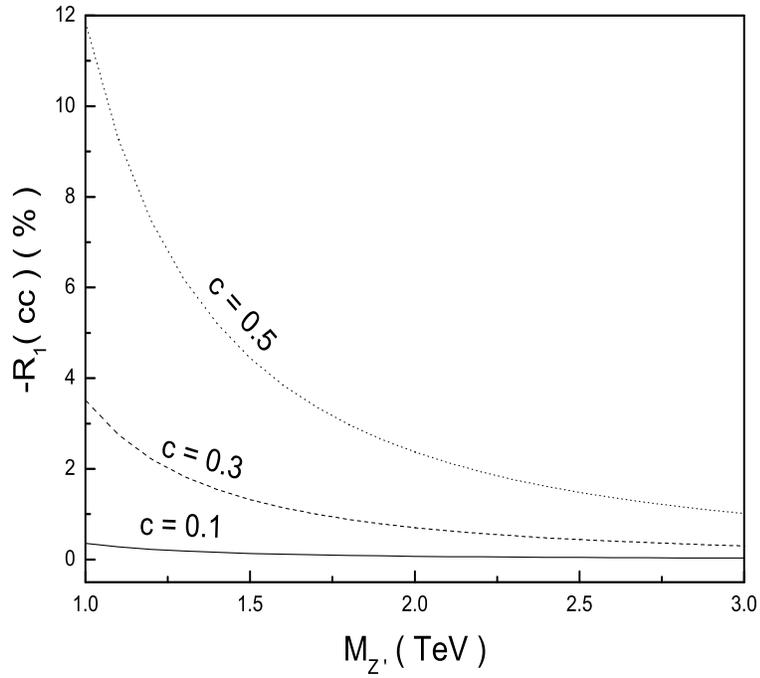,width=320pt,height=300pt} \vspace{-1cm}
\hspace{0.5cm} \caption{Same as Fig.3 but for the process
$e^{+}e^{-}\rightarrow \overline{c}c$.} \label{ee}
\end{center}
\end{figure}

The relative correction parameters $R_{2}(\overline{f}f)$ with
$f=l,\ b,\ c$ are plotted as functions of the $B'$ mass $M_{B'}$
for three values of the mixing parameter $c'$ in Fig.6--8.
Comparing these figures with Fig.3--5, we find that the
contributions of gauge boson $B'$ exchange to these processes are
larger than those of gauge boson $Z'$ exchange in wide range of
the parameter space. This is mainly because the heavy photon $B'$
is lighter than the gauge boson $Z'$ in most of the parameter
space of the LH model. However, the corrections of $B'$ exchange
to these processes may be positive or negative, which are
dependent on the value of the $B'$ mass $M_{B'}$. The peak of the
relative correction resonance emerges when the $B'$ mass $M_{B'}$
is approximately equal to $480GeV$ or $520GeV$ for the c. m.
energy $ \sqrt{s} = 500GeV$.

\begin{figure}[htb]
\vspace{0.5cm}
\begin{center}
\epsfig{file=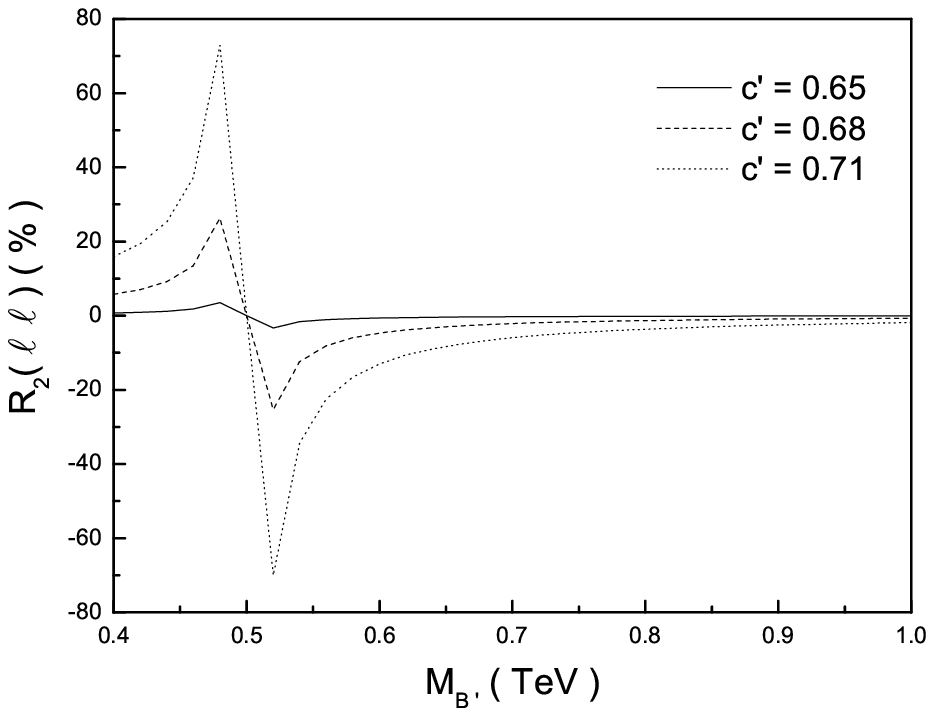,width=320pt,height=300pt} \vspace{0cm}
\hspace{0.5cm} \caption{The relative correction parameter $R_{2}$
for the process $e^{+}e^{-}\rightarrow \overline{l}l$ as function
\hspace*{1.8cm}of $M_{B'}$ for $c'=0.65$(solid line),
$c'=0.68$(dashed line) and $c'=0.71$(dotted \hspace*{1.8cm}line).}
\label{ee}
\end{center}
\end{figure}

\begin{figure}[htb]
\vspace{-0.5cm}
\begin{center}
\epsfig{file=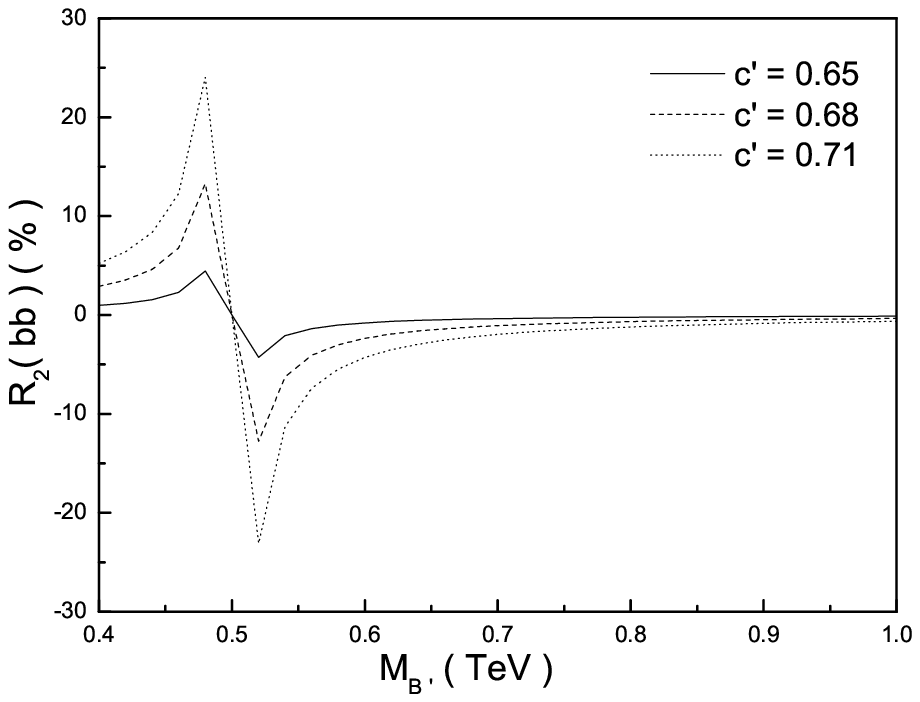,width=320pt,height=300pt} \vspace{-1cm}
\hspace{0.5cm} \caption{Same as Fig.6 but for the process
$e^{+}e^{-}\rightarrow \overline{b}b$.} \label{ee}
\end{center}
\end{figure}
\vspace*{9cm}
\begin{figure}[htb]
\vspace*{-11cm}
\begin{center}
\epsfig{file=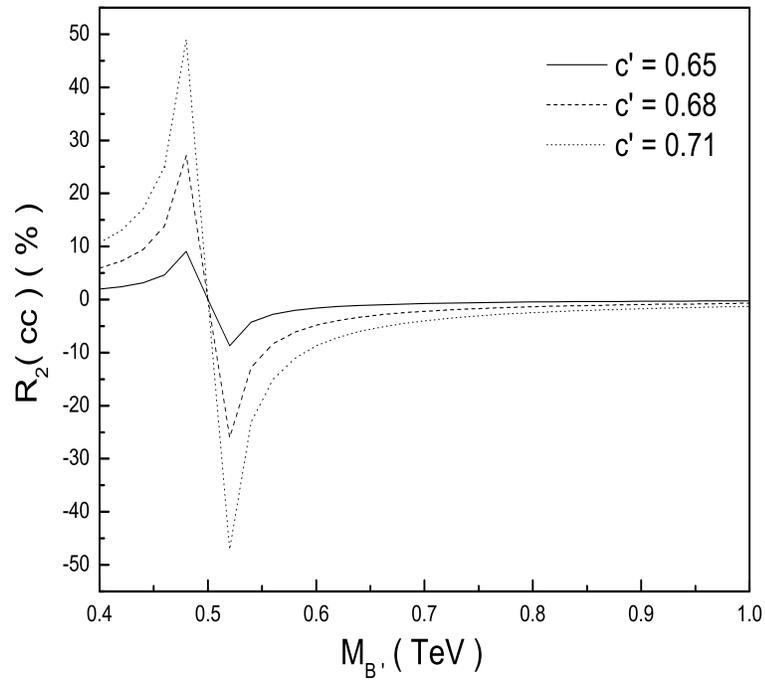,width=320pt,height=300pt} \vspace{-1cm}
\hspace{0.5cm} \caption{Same as Fig.6 but for the process
$e^{+}e^{-}\rightarrow \overline{c}c$.} \label{ee}
\end{center}
\end{figure}

From Figs.6--8, we can see that the contributions of the gauge
boson $B'$ to the processes
$e^{+}e^{-}\rightarrow\overline{\tau}\tau( \overline{\mu}\mu)$ are
larger than those to the processes
$e^{+}e^{-}\rightarrow\overline{b}b( \overline{c}c)$ and the gauge
boson $B'$ is most sensitive to the processes
$e^{+}e^{-}\rightarrow\overline{l}l$. For example, for
$M_{B'}=600GeV$ and $c'=0.71$, the absolute values of
$R_{2}(\overline{f}f)$ are $13\%$, $8.7\%$, and $4.3\%$ for
$e^{+}e^{-}\rightarrow\overline{l}l$,
$e^{+}e^{-}\rightarrow\overline{c}c$, and
$e^{+}e^{-}\rightarrow\overline{b}b$, respectively. However, if we
take $M_{B'}\geq750GeV$ and $0.62 \leq c'\leq 0.73$, the absolute
value of the relative correction parameter $R_{2}(\overline{l}l)$
is smaller than $5\%$. Thus, with reasonable values of the
parameters, we can detect the possible signals of the gauge boson
$B'$ via the processes $e^{+}e^{-}\rightarrow\overline{f}f$ in the
future LC experiments.

From Eqs.(13)-(17), we can see that the relative correction
resonance emerges when the $B'$ mass $M_{B'}$ approaches the c. m.
energy $ \sqrt{s} = 500GeV $ as shown in Figs.(6)-(8). The
resonance values of the relative correction parameter
$R_{2}(\overline{f}f)$ are strongly dependent on the coupling
strength of the gauge boson $B'$ with the light fermions. However,
one can see from Eqs.(6)-(8) that all gauge couplings of $B'$ with
light fermions vanish for $ c' = \sqrt{2/5}\approx 0.63 $(A
suitable value of the mixing parameter $c'$ can make the gauge
boson $B'$ not give too large contributions to some electroweak
observables.). Thus, we can use this feature to determine the
values of $M_{B'}$ and $c'$ in the future LC experiments.

\noindent{\bf IV. Probing limits of the new gauge bosons $Z'$ and
$B'$}

In this section, we discuss the realistic observability limits on
the free parameters of the new gauge bosons $Z'$ and $B'$, such as
$ M_{Z'}$, $ M_{B'}$, $c$, and $c'$,  by performing $x^{2}$
analysis, i.e. by comparing the deviations of the measured
observables from the SM predictions with the expected experimented
uncertainty including the statistical and the systematic one. From
our discussions in Sec.III, we can see that the total cross
sections $\sigma(\overline{f}f)$ of the processes
$e^{+}e^{-}\rightarrow\overline{f}f$ are rather sensitive to the
relevant free parameters of $Z'$ and $B'$. Thus, we will take this
observable as an example in this analysis. For the cross section
$\sigma(\overline{f}f)$, the $x^{2}$ function is defined as:
\begin{equation}
x^{2}=[\frac{\Delta\sigma(\overline{f}f)}{\delta\sigma(\overline{f}f)}]^{2},
\end{equation}
where $\delta\sigma(\overline{f}f)$ is the expected experimental
uncertainty about the cross section $\sigma(\overline{f}f)$
including both the statistical and systematic uncertainties at the
future LC experiments. The allowed values of the $Z'$ and $B'$
parameters by observation of the deviation
$\Delta\sigma(\overline{f}f)$ can be estimated by imposing
$x^{2}>x^{2}_{C.L.}$, where the actual value of $x^{2}_{C.L.}$
specifies the desired 'confidence' level. In the following
estimation, we will take $x^{2}_{C.L.}=3.84$ for $95\%$ C.L. and
for one parameter fit.

The square of the expected uncertainty about the cross section
$\sigma(\overline{f}f)$ have been given by Ref.[19], which can be
written as:
\begin{equation}
[\delta\sigma(\overline{f}f)]^{2}
=\frac{[\sigma(\overline{f}f)]^{2}}{{N_{tot}}^{exp}}
+[\sigma(\overline{f}f)]^{2}[\frac{p_{e}^{2}
p_{\overline{e}}^{2}}{D^{2}}(\varepsilon_{e}^{2}
+\varepsilon_{\overline{e}}^{2})+\varepsilon_{\pounds}^{2}],
\end{equation}
where $D=1-p_{e}p_{\overline{e}}$, in which $p_{e}$ and
$p_{\overline{e}}$ are the degrees of longitudinal electron and
positron polarization, respectively.
$N_{tot}^{exp}=N_{L,F}+N_{R,F}+N_{L,B}+N_{R,B}$ is the total
number events observed in the future LC experiment with polarized
beams,  which can also be represented as
$N_{tot}^{exp}=D\pounds_{int}\varepsilon\sigma(\overline{f}f)$.
The parameter $\varepsilon$ is the experimental efficiency for
detecting the final state fermions. In the following estimation,
we will take the commonly used reference values of these
parameters, $\varepsilon=95\%$ for $\overline{\mu}\mu$ or
$\overline{\tau}\tau$; $\varepsilon=60\%$ for $\overline{b}b$ and
$\varepsilon=35\%$ for $\overline{c}c$, $\varepsilon_{e}=\delta
p_{e}/p_{e}=0.5\%$, $\varepsilon_{\overline{e}}=\delta
p_{\overline{e}}/p_{\overline{e}}=0.5\%$, and
$\varepsilon_{\pounds}=\delta\pounds_{int}/\pounds_{int}=0.5\%$.

From above discussions, we can see that the $x^{2}$ function is
mainly dependent of the parameters $(M_{Z'}$, $c)$ and $(M_{B'}$,
$c')$ for the gauge bosons $Z'$ and $B'$, respectively. So, we can
use these equations to investigate the limits on the free
parameters of the gauge bosons $Z'$ and $B'$ in the cases of $Z'$
discovery and $B'$ discovery in the future LC experiments with
$\sqrt{s}=500GeV$ and $\pounds_{int}=340fb^{-1}$ [1] and give the
discovery upper bounds on $M_{Z'}$ and $M_{B'}$ for the fixed
values of the mixing parameters $c$ and $c'$. Our numerical
results for the processes $e^{+}e^{-}\rightarrow\overline{f}f$
with $f=l,\  b,$ or $c$ are summarized in Fig.9 and Fig.10, in
which we plot the discovery upper bounds on $M_{Z'}$ and $M_{B'}$
at $95\%$C.L. as function of the mixing parameters $c$ and $c'$,
respectively. We have assumed $p_{e}=0.8$ and
$p_{\overline{e}}=0.6$.

From Fig.9, one can see that the value of the discovery upper
bound on the $Z'$ mass $M_{Z'}$ increases as the mixing parameter
$c$ increasing. For $c < 0.3$, the allowed maximal value of
$M_{Z'}$ is approximately smaller than $1TeV $. Considering the
constraints of the precision measurement data on the LH model, the
mass $M_{Z'}$ of the gauge boson $Z'$ should be larger than
$1TeV$[11,12]. Thus, for $c < 0.3$, the gauge boson $Z'$ can not
be detected via the processes
$e^{+}e^{-}\rightarrow\overline{f}f(f=l,\ b,$ or $c$) in the
future LC experiment with $\sqrt{s}=500GeV$ and
$\pounds_{int}=340fb^{-1}$. However, for the large value of the
mixing parameter $c$, it is not this case. For example, for
$c=0.5$, the $Z'$ mass $M_{Z'}$ can be explored up to $1.9TeV$,
$2.8TeV$, and $2.2TeV$ via the processes
$e^{+}e^{-}\rightarrow\overline{l}l$,
$e^{+}e^{-}\rightarrow\overline{b}b$, and
$e^{+}e^{-}\rightarrow\overline{c}c$, respectively. From this
figure, we can also obtain the conclusion that the $Z'$ is most
sensitive to the processes $e^{+}e^{-}\rightarrow\overline{b}b$
and its virtual effects are most easy to be observed via this
process in the future LC experiments.

\begin{figure}[htb]
\vspace{-1cm}
\begin{center}
\epsfig{file=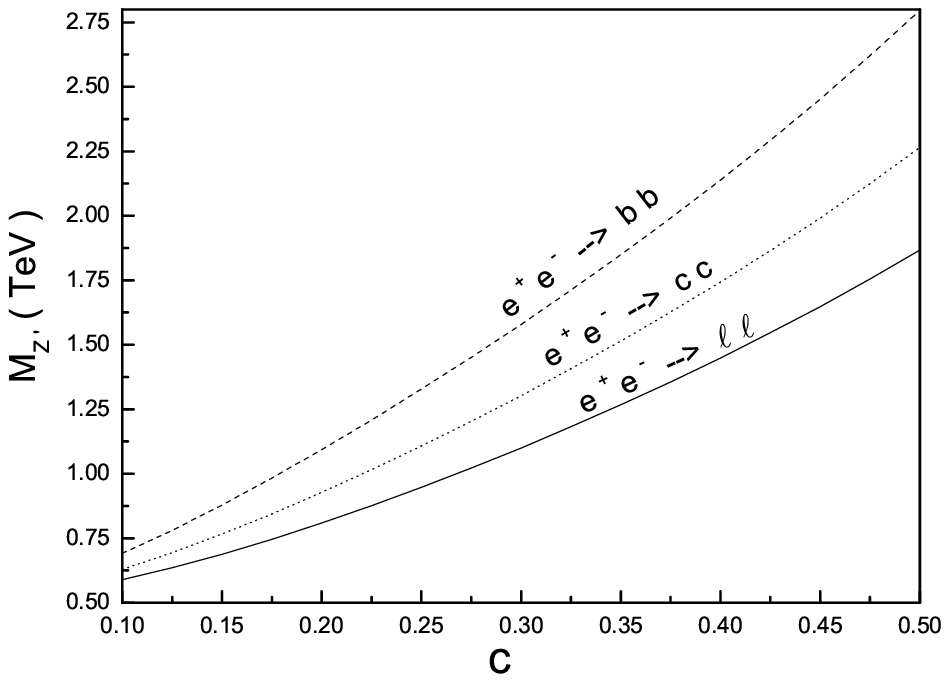,width=320pt,height=300pt} \vspace{-1.5cm}
\hspace{0.5cm} \caption{The $Z'$ mass $M_{Z'}$ as a function of
the parameter $c$ for $e^{+}e^{-}\rightarrow \overline{l}l$(solid
line), \hspace*{1.9cm}$e^{+}e^{-}\rightarrow \overline{b}b$(dashed
line) and $e^{+}e^{-}\rightarrow \overline{c}c$(dotted line).}
\label{ee}
\end{center}
\end{figure}

In general, as long as the $B'$ mass $M_{B'}$ is in the range of
$0.4TeV \sim 1TeV$, all of the processes
$e^{+}e^{-}\rightarrow\overline{f}f$($f=l,\ b$, and $c$) can be
used to detecting the possible signals of the neutral gauge boson
$B'$ in wide range of the parameter space of the LH model. The
gauge boson $B'$ is most sensitive to the processes
$e^{+}e^{-}\rightarrow\overline{l}l$. However, the electroweak
precision data give a severe constraint on the LH model and a
large portion of the parameter space has been ruled out. In the
parameter space, $0.62\leq c'\leq 0.73$, allowed by the
electroweak precision constraints, the discovery upper bounds of
$M_{B'}$ are largely reduced. From Fig.10 we can see that, for
$c'=0.65$, the $B'$ mass $M_{B'}$ can be explored only up to
$525GeV(564GeV)$ via the processes
$e^{+}e^{-}\rightarrow\overline{b}b(\overline{c}c$) in the future
LC experiment with $\sqrt{s}=500GeV$ and
$\pounds_{int}=340fb^{-1}$. But the $B'$ mass $M_{B'}$ can be
explored up to $1.26TeV $ via the processes
$e^{+}e^{-}\rightarrow\overline{l}l$ for $c'=0.73$. For the mixing
parameter $c'$ in the range of $0.64 \sim 0.73$, the $B'$ mass
$M_{B'}$ can be explored from $507GeV $ to $1.26TeV $ via the
processes $e^{+}e^{-}\rightarrow\overline{l}l$ in the future LC
experiments. Thus, we expect that, with reasonable values of the
free parameters of the LH model, the possible signals of the gauge
boson $B'$ can be observed in the future LC experiments.

\begin{figure}[htb]
\vspace{-0.5cm}
\begin{center}
\epsfig{file=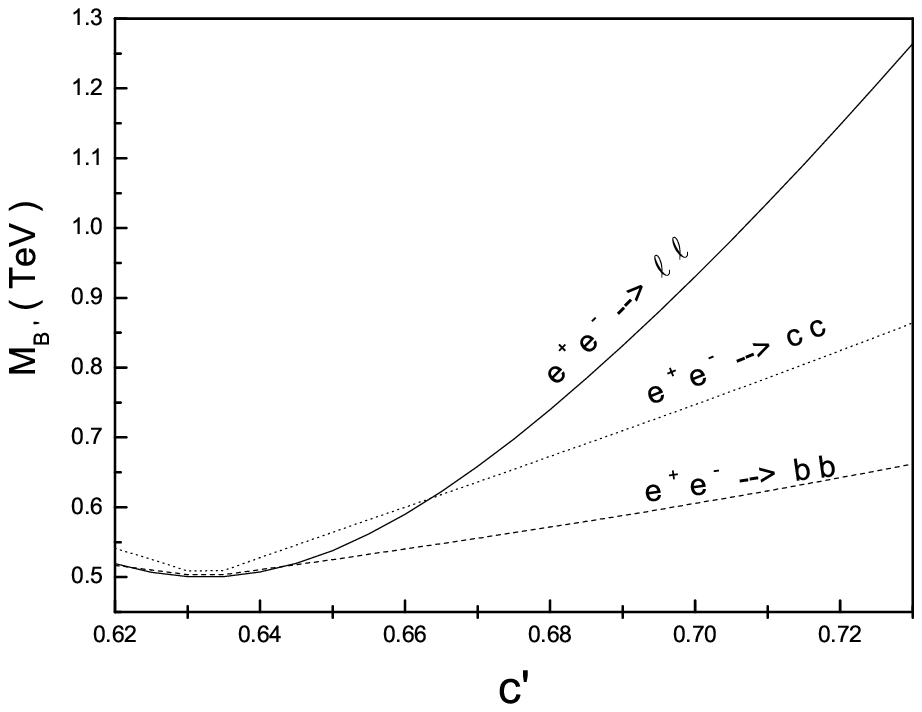,width=320pt,height=300pt} \vspace{-1cm}
\hspace{0.5cm} \caption{The $B'$ mass $M_{B'}$ as a function of
the parameter $c'$ for $e^{+}e^{-}\rightarrow \overline{l}l$(solid
line), \hspace*{2.0cm}$e^{+}e^{-}\rightarrow \overline{b}b$(dashed
line) and $e^{+}e^{-}\rightarrow \overline{c}c$(dotted line).}
\label{ee}
\end{center}
\end{figure}

\noindent{\bf V. Conclusions and discussions}

Little Higgs  models have generated much interest as possible
alternative to weak scale supersymmetry. The LH model is one of
the simplest and phenomenologically viable models, which realizes
the little Higgs idea. The LH model predicts the existence of the
new gauge bosons $Z'$ and $B'$. The possible signals of these new
particles might be detected in the future high energy experiments.

In this paper, we calculate the corrections of the gauge bosons
$Z'$ and $B'$ to the processes
$e^{+}e^{-}\rightarrow\overline{f}f$ with $f=\tau,\ \mu, \ b,$ or
$c$ and further discuss the possibility of detecting these new
particles via these processes in the future LC experiments with
$\sqrt{s}=500GeV$ and $\pounds_{int}=340fb^{-1}$. We find that the
gauge boson $Z'$ is most sensitive to the process
$e^{+}e^{-}\rightarrow\overline{b}b$, while the gauge boson $B'$
is most sensitive to the process
$e^{+}e^{-}\rightarrow\overline{l}l$. In wide range of the
parameter space of the LH model, the possible signals of the gauge
bosons $Z'$ and $B'$ can be detected via the processes
$e^{+}e^{-}\rightarrow\overline{f}f$ in the future LC experiments.
However, the LH model can produce significant contributions to
observables in most of the parameter space. Thus, the electroweak
precision data give a severe constraint on the LH model. A wide
range of the parameter space has been ruled out and the allowed
parameter space is $ f = 1\sim 2TeV $, $ 0 \leq c \leq 0.5 $, and
$ 0.62 \leq c' \leq 0.73 $[5,11,12,13,14]. In the allowed
parameter space, the possible signal of the gauge boson $Z'$ is
very difficult to be detected and it might be possible to detect
the gauge boson $Z'$ only for $ 0.4 \leq c \leq 0.5 $ and $
M_{Z'}\leq 1.25 TeV $. Taking into account the electroweak
constraints on the free parameters, the gauge boson $B'$ might be
observable via all of the processes
$e^{+}e^{-}\rightarrow\overline{f}f$ for $ M_{B'}\leq 750 GeV $
and $ c'\geq 0.65$.

 In the parameter space consistent with the electroweak precision
constraints, we further discuss the discovery upper bounds on the
$Z'$ mass $M_{Z'}$ and the $B'$ mass $M_{B'}$ via the processes
$e^{+}e^{-}\rightarrow\overline{f}f$ in the future LC experiments
with $\sqrt{s}=500GeV$ and $\pounds_{int}=340fb^{-1}$. We find
that, for $ 0.3 \leq c \leq 0.5 $, the $Z'$ mass $M_{Z'}$ can be
explored up to $ 2.8TeV $ via the process
$e^{+}e^{-}\rightarrow\overline{b}b$.  For the mixing parameter
$c'$ in the range of $0.64 \sim 0.73$, the $B'$ mass $M_{B'}$ can
be explored from $507GeV $ to $1.26TeV $ via the processes
$e^{+}e^{-}\rightarrow\overline{l}l$.

Certainly, the modification to the relation between the SM free
parameters and the correction terms to the tree-level
$Z\overline{f}f$ couplings can also produce corrections to the
processes $e^{+}e^{-}\rightarrow\overline{f}f$, which might mix
with the corrections arising from $Z'$ exchange or $B'$ exchange.
However, our calculation results show that these two kinds of
contributions are all smaller than those of $Z'$ exchange or $B'$
exchange at least by two orders of magnitude in wide range of the
parameter space of the LH model. Thus, comparing with the direct
contributions of $Z'$ and $B'$, the contributions from the
modification to the relation between the SM free parameters and
from the tree-level correction terms can be safely neglected.

In conclusion, in a large portion of the parameter space
consistent with the electroweak precision constraints, the new
gauge boson $B'$ might be detected via the processes
$e^{+}e^{-}\rightarrow\overline{f}f$ in the future LC experiments.
The gauge boson $Z'$ can only be detected in small part of the
parameter space. However, observation of such gauge bosons will
not prove that they are the new particles predicted by the LH
model. It is need to discuss the possible decay channels and
possible characteristic signatures, which have been extensively
studied in Ref.[20].

\vspace{0.5cm} \noindent{\bf Acknowledgments}

C. X. Yue would like to thank the ICTP( Trieste, Italy) for
hospitable and the stimulating environment when the revised
version of the paper was finished. This work was supported in part
by the National Natural Science Foundation of China under the
grant No.90203005 and No.10475037 and the Natural Science
Foundation of the Liaoning Scientific Committee(20032101).

\null

\newpage

\begin{thebibliography}{99}
\bibitem{1}
        T. Abe et al., [American Linear Collider Working Group], {\em hep-ex}/{\bf0106057};
        J. A. Aguilar-Saavedra et al.
        [{\it ECFA/DESY } Physics Working Group], {\em hep-ph}/{\bf
        0106315}; K. Abe et al., [{\it ACFA }Linear Collider Working Group], {\em hep-ph}/{\bf 0109166}.
\bibitem{2}
        N. Arkani-Hamed, A. G. Cohen, E. Katz, A. E.
        Nelson, {\bf JHEP 0207}(2002)034.
\bibitem{3}
        N. Arkani-Hamed, A. G. Cohen and H. Georgi,
         {\em  Phys. Lett. B}{\bf 513}(2001)232; N. Arkani-Hamed, A. G. Cohen,
         T. Gregoire and J. G. Wacker,
         {\bf JHEP 0208}(2002)020; N. Arkani-Hamed, A. G. Cohen,
         E. Katz, A. E.  Nelson, T. Gregoire and J. G. Wacker, {\bf JHEP
         0208}(2002)021; I. Low, W. Skiba and
         D. Smith, {\em Phys. Rev. D}{\bf 66}(2002)072001; M.
         Schmaltz, {\em Nucl. Phys. Proc. Suppl.} {\bf
         117}(2003)40; D. E. Kaplan and M. Schmaltz,
         {\bf JHEP 0310}(2003)039.
\bibitem{4}
         J. G. Wacker, {\em hep-ph}/{\bf0208235}; S. Chang and J. G. Wacker,
         {\em Phys. Rev. D}{\bf69}(2004)035002;
         W. Skiba and J. Terning, {\em Phys. Rev. D}{\bf 68}(2003)075001; S.
         Chang, {\bf JHEP 0312}(2003)057; M. Schmaltz, {\bf JHEP 0408}(2004)056.
\bibitem{5}
         T. Han, H. E. Logan, B. McElrath and L. T. Wang,
        {\em Phys. Rev. D}{\bf 67}(2003)095004.
\bibitem{6}
         M. Perelstein, M. E. Peskin and A. Pierce, {\it Phys. Rev. D}{\bf 69}(2004)075002.
\bibitem{7}
        G. Burdman, M. Perelstein and A. Pierce, {\it Phys. Rev. Lett.} {\bf
        90} (2003) 241802.
\bibitem{8}
         S. C. Park and J. Song, {\it Phys. Rev. D}{\bf 69}(2004)115010.
\bibitem{9}
        T. Han. H. E. Logan, B. McElrath and L. T. Wang, {\it Phys. Lett. B}{\bf 563}(2003)191;
         H. E. Logan, {\it Phys. Rev. D}{\bf70}(2004)115003; G. A. Gonzalez-Sprinberg,
         R. Martinez,
         and J. Alexis Rodriguez, {\it hep-ph}/{\bf 0406178}; Gi-Chol Cho and Aya Omote,
         {\it Phys. Rev. D}{\bf 70}(2004)057701; Jaeyong Lee, {\bf JHEP 0412}(2004)065.
\bibitem{10}
         Chong-Xing Yue, Shun-Zhi Wang, Dong-Qi Yu, {\it Phys. Rev. D}{\bf 68}(2003)115004.
\bibitem{11}J. L. Hewett, F. J. Petriello and T. G. Rizzo, {\bf JHEP 0310}(2003)062.
\bibitem{12}C. Csaki, J.Hubisz, G. D. Kribs, P. Meade, and J. Terning, {\it Phys. Rev. D}{\bf
         67}(2003)115002.
\bibitem{13}C. Csaki, J.Hubisz, G. D. Kribs, P. Meade, and J. Terning, {\it Phys. Rev. D}{\bf
         68}(2003)035009.
\bibitem{14}
        T. Gregoire, D. R. Smith and J. G. Wacker, {\it Phys. Rev. D}{\bf
         69}(2004)115008
\bibitem{15}
         R. S. Chivukula, N. Evans and E. H. Simmons, {\it Phys. Rev. D}{\bf
         66}(2002)035008; S. Chang and Hong-Jian He, {\it Phys. Lett. B}{\bf
         586}(2004)95; I. Low, {\bf JHEP 0410}(2004)067.
\bibitem{16}
         Mu-Chun Chen and S. Dawson, {\it Phys. Rev. D}{\bf
         70}(2004)015003; R. Casalbuoni, A. Deandrea, M. Oertel,
         {\bf JHEP 0402}(2004)032; Chong-Xing Yue and Wei Wang, {\it Nucl. Phys. B}
         {\bf 683}(2004)48.
\bibitem{17}
         W. Kilian and J. Reuter, {\it Phys. Rev. D}{\bf70}(2004)015004.
\bibitem{18}
        A. Djouadi, A. Leike, T. Riemann, D. Schaile, C. Verzegnassi, {\it Z. Phys. C}
        {\bf 56}(1992)289;
         S. Godfrey, {\it Phys. Rev. D}{\bf 51}(1995)1402; S. Riemann, {\it hep-ph}/
         {\bf 9610513};
         A. Leike, S. Riemann, {\it Z. Phys. C}{\bf 75}(1997)341;
        S. Godfrey, J. L. Hewett, L. E. Price, {\it hep-ph}/{\bf 9704291}; A.
        Leike, {\it Phys. Rept.} {\bf 317}(1999)143; R. Casalbuoni, S.
        De Curtis, D. Dominici, R. Gatto, S. Riemann,
        {\it hep-ph}/{\bf 0001215}.
\bibitem{19}
         A. A. Babich, et al., {\it Phys. Lett. B} {\bf518}(2001)128;
         A. A. Pankov and N. Paver, {\it Eur. Phys. J. C}{\bf 29}(2003)313;
         A. V. Tsytrinov and A. A. Pankov, {\it hep-ph}/{\bf 0401020}.
\bibitem{20}
         G. Azuelos, et al., {\it hep-ph}/{\bf 0402037}.
\end{thebibliography}
\end{document}